\documentclass[aps,pra,twocolumn,floatfix,longbibliography]{revtex4-1}
\usepackage[stretch=10]{microtype}
\usepackage{times}
\usepackage{amsfonts,amsmath,amssymb,amsthm,amscd}
\usepackage{acronym,array,bm,color,dcolumn}
\usepackage[pdftex]{graphicx}
\usepackage{nomencl,revsymb,upgreek,url}
\usepackage{hyperref}
\hypersetup{colorlinks=true, pdfauthor=Kevin C. Young, citecolor=cyan, urlcolor=blue}

\newcommand{\bra}[1]{\ensuremath{\left\langle{#1}\right\vert}}
\newcommand{\ket}[1]{\ensuremath{\left\vert{#1}\right\rangle}}

\newcommand{\abs}[1]{\left\vert #1 \right\vert}
\newcommand{\expect}[1]{\ensuremath{\left\langle{#1}\right\rangle}}

\newcommand{\erf}[1]{Eq.~(\ref{#1})}

\newcommand{\haqc}{\ensuremath{H_{\rm AQC}(t)}}

\newcommand{\haqcb}{\bar H_{\rm {AQC}}(t)}
\newcommand{\hc}{\ensuremath{H_{\rm C}}}
\newcommand{\hb}{\ensuremath{H_{\rm B}}}
\newcommand{\hr}{\ensuremath{H_{\rm S-R}}}
\newcommand{\uc}{\ensuremath{\mathcal{U}_{\rm C}}}

\newcommand{\hddt}{\tilde H_{\rm DD}(t)}

\newcommand{\proj}{\mathbf{P}}
\newcommand{\projzero}{\mathbf{P}_{\!\mathbf{0}}}

\newcommand{\hilbert}{\mathcal{H}}
 \newcommand{\nth}{\textsuperscript{th}}

\newcommand{\bnu}{{\boldsymbol{\nu}}}

\newcommand{\enu}{E_{\boldsymbol{\nu}}}
\newcommand{\Sj}{S_{\mathbf j}}
\newcommand{\bj}{\mathbf{j}}
\newcommand{\bk}{\mathbf{k}}
\newcommand{\bz}{\mathbf{0}}

\newcommand{\prsec}[1]{\section{#1}}
\newcommand{\prsubsec}[1]{\subsection{#1}}

\begin{document}

\title{Error suppression and error correction in adiabatic quantum computation I: \\techniques and challenges}

\author{Kevin C.~Young}
\email[Electronic address: ]{kyoung@sandia.gov}
\affiliation{Scalable \& Secure Systems Research (08961),
Sandia National Laboratories, Livermore, CA 94550}
\author{Mohan Sarovar}
\affiliation{Scalable \& Secure Systems Research (08961),
Sandia National Laboratories, Livermore, CA 94550}
\author{Robin Blume-Kohout}
\affiliation{Advanced Device Technologies (01425),
Sandia National Laboratories, Albuquerque, NM 87185}

\date{\today}

\begin{abstract}
\notag \noindent
Adiabatic quantum computation (AQC) has been lauded for its inherent robustness to control imperfections and relaxation effects.  A considerable body of previous work, however, has shown AQC to be acutely sensitive to noise that causes excitations from the adiabatically evolving ground state.  In this paper, we develop techniques to mitigate such noise, and then we point out and analyze some obstacles to further progress.  First, we examine two known techniques that leverage quantum error-detecting codes to suppress noise, and show that they are intimately related and may be analyzed within the same formalism.  Next, we analyze the effectiveness of such error suppression techniques in AQC, identify critical constraints on their performance, and conclude that large-scale, fault tolerant AQC will require error \emph{correction}, not merely suppression.  Finally, we study the consequences of encoding AQC in quantum stabilizer codes, and discover that generic AQC problem Hamiltonians rapidly convert physical errors into uncorrectable logical errors.  We present several techniques to remedy this problem, but all of them require unphysical resources, suggesting that the adiabatic model of quantum computation may be fundamentally incompatible with stabilizer quantum error correction.
%
\end{abstract}
\maketitle


\noindent Adiabatic quantum computation (AQC) is expected to be inherently robust against certain errors, such as dephasing and energy relaxation\cite{Farhi:2000tw,Childs2001}.  This robustness suggests the possibility of an easier route to scalable quantum computation than the conventional gate-based ``circuit'' model, with less stringent requirements for fault tolerance and fewer resources devoted to error suppression and correction.  However, AQC's inherent robustness is not sufficient for fault tolerance.  For example, several studies \cite{Childs2001,Gai-2006,Tie.Sch-2007,Ami.Tru.etal-2009,Veg.Ban.etal-2010} show that single-qubit noise can drive undesirable transitions out of the adiabatic ground state.  In response, error \emph{suppression} techniques have been developed that can reduce the rate at which these transitions occur.  However, it is well understood that error suppression alone is insufficient for fault tolerance in the circuit model.  Fault tolerance requires an additional mechanism to remove the entropy generated by errors that do occur in the encoded system -- i.e., error \emph{correction}.  Since this thermodynamic argument is independent of the computational model, we reasonably expect that achieving fault-tolerant AQC will also require some form of error correction.  

In this paper, we address both error suppression and error correction in AQC, and prove several facts about them.  We begin our discussion in Section I with an overview of the failure mechanisms present in AQC. In Section II we discuss error suppression techniques based on error-detecting quantum stabilizer codes.  Currently known suppression strategies include: energy gap protection (EGP) \cite{Jordan:2006jb}, in which the addition of the stabilizer generators to the system Hamiltonian causes errors to incur large energetic penalties; dynamical decoupling (DD)\cite{Lid-2008}, whereby stabilizer generators are applied periodically as unitary operators, refocusing errors much like traditional spin echos; and Zeno effect suppression \cite{PazSilva:2012kp}, which prevents errors from accumulating through frequent measurements of the stabilizer generators.  These three techniques apparently operate by very different physical mechanisms.  However, Facchi \textit{et al.} \cite{Fac.Lid.etal-2004} have shown that both Zeno suppression and DD may be viewed as limiting cases of a more general mathematical framework.  We extend this unification by showing that DD and EGP may both be understood using the same formalism, and that the two methods are effectively equivalent in their error suppression power.  Our result shows that all three error suppression techniques may be considered functionally equivalent for AQC, with some important caveats related to their physical implementation and their behavior in the presence of a thermal bath.

With Section III, we turn our attention to error correction, where we discover two fundamental obstacles to adaptation of stabilizer code quantum error correction to AQC: (i) in the presence of the adiabatic Hamiltonian, correctable errors rapidly become uncorrectable errors and (ii) the resources required by circuit-model error correction, such as stabilizer measurements and unitary gates, are generally outside the scope of the adiabatic paradigm.  In Section IV of the paper, we explore possible ways to overcome these obstacles when adapting stabilizer-based error correction to AQC.  We identify alternative implementations of logical operators and continuous-time error correction by cooling as possible yet \emph{very} challenging routes forward.  Section V finishes with a discussion of our view of the future of adiabatic fault tolerance.  

The major results discussed in this manuscript are:
\begin{enumerate}
	\item In the Hamiltonian formalism, the energy gap protection and dynamical decoupling methods for error 
		  suppression based on stabilizer encodings may be described by a unified mathematical formalism.
	\item AQC appears to be fundamentally incompatible with stabilizer quantum error correction. Patching this incompatibility requires unphysical resources.
\end{enumerate}

A companion paper entitled, ``\textit{Error suppression and correction in adiabatic quantum computation: non-equilibrium dynamics}" \cite{Sarovar:2013}, develops a dynamical model for describing error suppression and correction in AQC, and discusses most of the results in this paper from a dynamical perspective. 

\prsec{The Quantum Adiabatic Algorithm \hspace{10cm}and Important Failure Modes}
\label{sec:aqc_fail}
\noindent
The quantum adiabatic algorithm \cite{Farhi:2000tw} operates by slowly changing an $N$-qubit system's Hamiltonian from a simple separable Hamiltonian $H_{\rm init}$, whose ground state is easily prepared, to a final \emph{target} Hamiltonian, $H_{\rm prob}$, whose highly nontrivial ground state encodes the solution to a problem of interest.  The adiabatic theorem promises that, if the Hamiltonian is changed slowly enough, then the system will remain in its (time-varying) ground state, and thus the solution can be read out by measuring the final state.  In the simplest case, the closed-system dynamics may be described by a time-dependent Hamiltonian,
\begin{equation}
\label{eq:haqc}
	H_{\rm AQC}(t) = (1-s(t)) H_{\rm init} + s(t) H_{\rm prob},
\end{equation}
where $s(0)=0$, $s(T) = 1$, and $T$ is the adiabatic interpolation time.  More complicated interpolation schemes, including the addition of ancillary Hamiltonians, have been considered elsewhere\cite{Roland2002,Rezakhani2009}, but are unnecessary for our discussions here.  Importantly, the adiabatic model does not require the ability to perform high-quality quantum gates or measurements during the computation, key elements of fault tolerant circuit-model quantum computation.

The adiabatic model is expected to be robust to some errors that plague other computation models, such as the cluster-state or circuit models.   In particular, Ref.~\cite{Childs2001} showed that AQC possesses an inherent robustness to both control errors and some forms of decoherence.  In this section we give a brief overview what \emph{can} go wrong with adiabatic computations and discuss which failure modes can be suppressed with current techniques.

Perhaps the best known failure mode of AQC is diabatic errors (a.k.a. Landau-Zener transitions), in which the Hamiltonian is varied too quickly and the state fails to track the instantaneous ground state.  The obvious solution to this problem is to perform the interpolation more slowly, although identifying diabatic errors and determining the maximum allowable speed are nontrivial.  For certain problems, the location of the minimum energy gap between the ground and first excited state is well known (e.g., for adiabatic Grover search, it happens exactly in the middle\cite{Roland2002}).  This knowledge permits efficiently varying the interpolation speed so that it proceeds rapidly in regions where the gap is large, and slowly in regions where the gap is small.  Unfortunately, the location and magnitude of the minimum gap are not known for most problems.  The blunt approach of slowing the entire interpolation is problematic because even simple problems may have exponentially small minimum gaps, leading to exponentially long interpolation times.  This problem is not the focus of this paper, but solving it will necessarily involve (or enable) great leaps in our understanding of the computational complexity of the adiabatic algorithm.

While the system can be quite susceptible to diabatic transitions, AQC is known to possess some intrinsic robustness to Hamiltonian control errors \cite{Childs2001}.  As long as the evolution remains adiabatic,
small perturbations to the intermediate Hamiltonians are likely to be unimportant.  However, errors in the \emph{final} Hamiltonian can be fatal, since it is the final Hamiltonian that encodes the problem to be solved.  In particular, if the final Hamiltonian is close to a critical point, small perturbations may drastically alter the character of the ground state.  This failure mode is addressed in \cite{Young2013c}.  

Finally, because the success of an adiabatic interpolation relies ultimately on the population in the final ground state, it is robust to environmental couplings that cause decoherence in the eigenbasis of $H(s)$ (sometimes called \emph{dephasing}). 
However, most system-bath couplings will
cause transitions from the adiabatic ground state as well as Lamb shifts of the system Hamiltonian.  The resulting open system dynamics may be radically different from those of the ideal closed system.   This paper is largely concerned with these errors, and we focus on techniques designed to suppress and correct the influence of these system-bath couplings in a manner consistent with the adiabatic paradigm.  Specifically, we attempt to avoid reliance on quantum gates and measurements as much as possible.  

\prsec{Suppressing errors in AQC} 
\label{sec:protecting}
\noindent
In this section we introduce quantum stabilizer codes and discuss their role in protecting adiabatic quantum computations.  A system undergoing adiabatic quantum evolution while coupled to an external environment/bath is described in a tensor-product Hilbert space, $\mathcal{H}_\textrm{sys} \otimes \mathcal{H}_\textrm{env}$, by a Hamiltonian
\begin{equation}
\label{eq:aqcham}
	H(t) =  \haqc{} + \sum_{j=1}^{n_e} E_j\otimes B_j  + \hb.
\end{equation}
Here $H_{\rm AQC}$ acts on the system and performs the adiabatic evolution, and $\hb$ is the bath Hamiltonian. $B_j$ is a bath operator, and $E_j$ is a single qubit Pauli error operator. The total number of system-bath coupling operators is $n_e$, and is generally proportional to the total number of qubits in the system.  Eigenstates of $H_{\rm AQC}(t)$ may be labeled as $\ket{n,k}_t$ according to their principal quantum number $n$ and an index $k$ distinguishing any degeneracy; the subscript labeling the time is necessary because the Hamiltonian is time-dependent (so $\ket{n,k}_t \ne \ket{n,k}_{t^\prime}$).  The system-bath interaction terms in the Hamiltonian can cause the computation to fail by inducing transitions out of the adiabatically evolving ground state.

\prsubsec{Quantum stabilizer codes}
\label{sec:stabilizers}
\noindent
All currently known techniques for suppressing the errors induced by the system-bath interaction terms rely on encoding the system in an error detecting stabilizer code \cite{Got-1997, Gaitan2008}.  ``Encoding'' comprises:
\begin{enumerate}
\item introducing (many) extra physical qubits to the system, and
\item mapping the original computational qubits (on which the computation is performed) into \emph{logical} qubits that are distributed across many physical qubits, much as in a classical repetition code.
\end{enumerate}
All the physical qubits together define a large system with a Hilbert space $\bar{\mathcal{H}}_{\rm sys}$.  The logical qubits are a subsystem, corresponding to a factor space $\mathcal{L}$, so the entire Hilbert space factors as $\bar{\mathcal{H}}_{\rm sys} = \mathcal{L}\otimes\mathcal{S}$.  The complementary subsystem $\mathcal{S}$ is the \emph{syndrome} subsystem, 
This factorization into subsystems is carefully chosen so that physical errors (on a small number of physical qubits) can  be detected by Pauli measurements on the syndrome.  These measurement operators, the \emph{stabilizer generators}, are the quantum analogue of parity checks in classical linear block codes.  The stabilizer generators generate an abelian subgroup of the Pauli group (the stabilizer group, comprising all possible products of generators), and are used to compactly define the code.  Stabilizers can be measured without disturbing the encoded quantum information, because logical qubit operators (by definition) commute with the stabilizers.

An encoded system is always initialized in a known eigenstate of all the stabilizers.  Subsequently measuring the stabilizers will reveal (detect) any error operation that anticommutes with one or more stabilizer generator (since such a \emph{detectable error} necessarily flips the sign of some stabilizer eigenvalues). These errors will be generated by terms in the system-bath Hamiltonian, whose effect on the system is to apply $E_j$ operations (\erf{eq:aqcham}) on the system. Thus we want to choose a code such that each $E_j$ in the system-bath interaction anticommutes with at least one of the stabilizer generators.  Utilizing a code with $N_g$ stabilizer generators adds $N_g$ physical qubits, enlarging the system's Hilbert space by a factor of $2^{N_g}$.  The encoding process replaces the original problem Hamiltonian $H_{\rm AQC}$ with an encoded Hamiltonian, in which the Pauli operators, $\sigma_x, \sigma_y, \sigma_z$ (that acted on physical bits in the unencoded Hamiltonian) are replaced by the code's logical operators, $\bar X, \bar Y, \bar Z$.  In addition, a time-dependent system control Hamiltonian, $\hc$, expressible in terms of the code's stabilizer generators, is added to implement any desired error suppression.  The encoded Hamiltonian is then:
\begin{equation}
\label{eq:aqchambar}
	\bar H(t) =  \haqcb + \hc(t) + \sum_{j=1}^{N_e} E_j\otimes B_j + \hb 
\end{equation}
We have assumed that the system-bath interaction remains qualitatively the same after the encoding, but is extended to $N_e > n_e$ terms to describe the extra qubits.  Importantly, we have assumed that any controls we apply act only on the system and have no effect on system-bath couplings.

States of the encoded system may now be labeled by the same two quantum numbers as before, but with $N_g$ additional binary quantum numbers, collected into the vector $\bnu$, indicating the eigenvalues of the stabilizer generators, $S_j\ket{n,k;\bnu}_{\!t} = (-1)^{\nu_j}\ket{n,k;\bnu}_{\!t}$, where $S_j\in \mathcal{S}$ is a generator of the stabilizer group and $\nu_j = \{0,1\}$ is the $j\nth$ element of the vector $\bnu$.  We shall refer to the subspace on which all the stabilizer eigenvalues are $+1$ (and therefore all $\nu_j=0$) as the \emph{codespace}.  States in the codespace will therefore be labeled as $\ket{n,k;\mathbf{0}}_t$.  The projector onto the codespace is time-invariant, and expressible in terms of the stabilizer generators as
\begin{equation}
\label{eq:projector}
	\projzero = \frac{1}{2^{N_g}}\prod_{k=1}^{N_g} (1+S_k).
\end{equation}
Encoding will help to protect logical information (stored initially in the codespace) by permitting active suppression of errors that cause transitions \emph{out} of the codespace.  Errors that mix states \emph{within} the codespace are necessarily high weight
\footnote{We define the weight of Pauli operator 
			to be the number of non-identity terms in the tensor 
			product.  So, for example, $XXIIIZ$ has weight 3, denoted $\mathbf{w}(XXIIIZ)=3$.} 
 (and therefore, hopefully, unlikely).  Encoding also makes it possible in principle to \emph{correct} errors, by using the results of stabilizer measurements to detect and identify errors, then inverting them.  However, such correction operations traditionally require resources (measurements and gates) that we have abjured.

In the following discussion we will make extensive use of the toggling frame, a rotating frame of reference defined in terms of the control and bath Hamiltonians.  Transformations to and from this frame are effected by the unitary operator,
\begin{equation}
\label{eq:Ucontrol}
	\uc(t_1,t_2) = \textrm{exp}\left(-i\int_{t_1}^{t_2} \left(\hc(s) + \hb\right) ds \right)
\end{equation}
There is no need to time order the integral because $\hc(s)$ may be written entirely in terms of the stabilizer generators of the code, all of which are mutually commuting, and $\hb$ acts on a different part of the Hilbert space.  The following notation is used for an operator $A$ in the toggling frame: $\tilde{A}(t) \equiv \uc^\dagger(t_1,t_2) A~ \uc(t_1,t_2)$, while states in the toggling frame are related to states in the Schrodinger picture by: $\vert\tilde\psi\rangle_t = \uc^\dagger(0,t) \ket{\psi}_t$. Evolution of states in this frame is generated by the toggling frame Hamiltonian: $\tilde{H}(t) \equiv\uc^\dagger(0,t) \left(\bar H(t) -\hc -\hb \right)\uc(0,t)$.

\prsubsec{Dynamical decoupling}
\label{sub:dynamical_decoupling}
\noindent
Dynamical decoupling (DD) is a well-known quantum control technique for suppressing errors produced by spurious terms in a system's Hamiltonian \cite{Viola1999}.  The methods were first applied to AQC in Ref.~\cite{Lid-2008}, with higher-order strategies shown to be particularly effective in Ref.~\cite{Quiroz2012}.  In DD, the stabilizer generators are applied as unitary operations by manipulating the control Hamiltonian $\hc$.  The sequence in which they are applied is given by the vector $\mathbf{n}$, at times given by $K(t)\in \mathbb{Z}$, so that at time $t$, the last operator applied to the system was $S_{\mathbf{n}_{K(t)}}$.  The unitary operator defining the toggling frame may then be written as,
$\uc^{\rm DD}(t) = \prod_{j=0}^{K(t)} S_{\mathbf{n}_j}$.  The operator $\uc(t)$ is an element of the full stabilizer group and therefore commutes with $\haqcb$.  In the toggling frame the Hamiltonian takes the form:
\begin{align}
	\hddt
		 \notag
		&= \bar H_{\rm AQC}(t) + \sum_{j=1}^{N_e} \tilde E_j^{\rm DD}(t) \otimes B_j(t)
\end{align}
Since $E_j$ is a Pauli operator, it either commutes or anti-commutes with each member of the stabilizer group and we may write, 
\begin{equation}
\label{eq:ejdd}
	\tilde E_j^{\rm DD}(t) = \uc^{\rm DD \dagger}(t) E_j \uc^{\rm DD}(t) = (-1)^{p_j(t)} E_j,
\end{equation}
where $p_j(t) = 0$ if $[E_j,\uc^{\rm DD}(t)]=0$ and $p_j(t) = 1$ if $\{E_j,\uc^{\rm DD}(t)\}=0$. A well-chosen DD sequence will cause $p_j(t)$ to rapidly alternate between $0$ and $1$, which modulates the the system-environment coupling (in the toggling frame) by a rapidly oscillating function of $t$. The unitary operator governing the evolution at time $t$  is $U(t,0) = \exp_+\{ -i\int_0^t ds \tilde{H}_\textrm{DD}(s)\}$, and if the DD sequence is well-chosen, the system-environment coupling averages to zero on timescales longer than the DD interpulse period (i.e., the integral vanishes thanks to the modulation factor $(-1)^{p(s)}$ in the exponential).

\begin{figure}[ht]
	\includegraphics[width=\columnwidth]{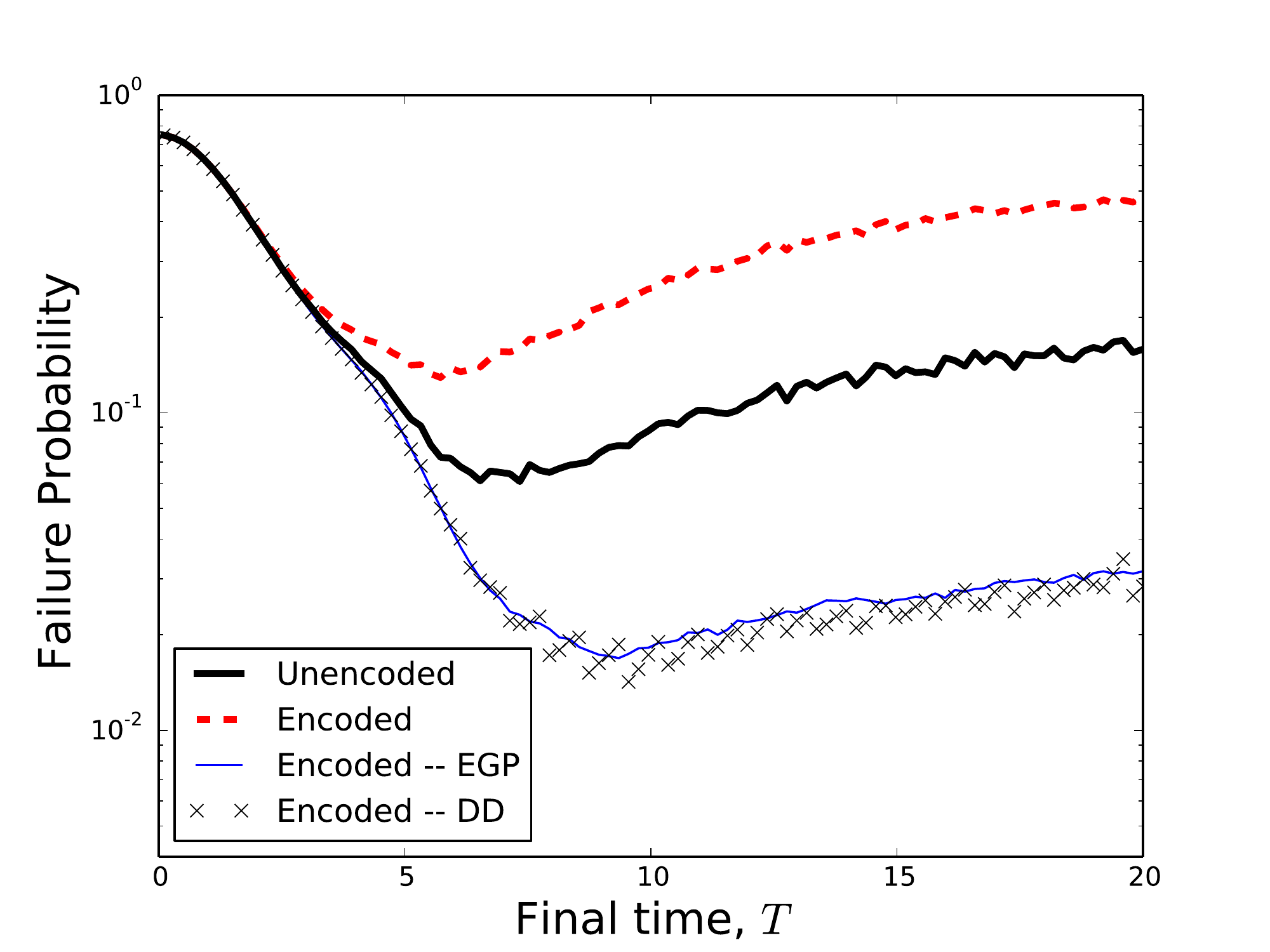}
	\caption{(color online) Logarithmic plot of the failure probabilities as a function of the total evolution time, $T$, for a two-logical-qubit AQC defined by the Hamiltonian, $H(t) = (1-t/T)\left(\sum_i X_i\right) + (t/T)\left(Z_1 + Z_2 + Z_1 Z_2\right)$ under the influence of Hamiltonian noise, $H_\eta(t)=\sum_i \eta_i(t)Z_i$.  In these numerical simulations $\eta_i$ is a classical stochastic process of $1/f$ type with spectrum $S(\omega)=10^{-3}/\omega$, and the success probabilities were averaged over 1000 instances of this noise process for each evolution time $T$.  Four simulations were performed: (i) the unencoded system (thick black line), (ii) the system encoded into four physical qubits using the [[4,2,2]] quantum code \cite{Got-1997} (red dashed line), (iii) the encoded system with EGP applied at strength $\Omega=1$ (thin blue line), (iv)  the encoded system with DD applied at frequency $\Omega/2\pi$ (black x's).  Initial performance increase is due to reduction of diabatic errors, while later performance degradation is due to the accumulation of errors.  Both EGP and DD are capable of suppressing this type of noise equally well, as shown by approximately equal success probabilities of cases (iii) and (iv). Encoding without error suppression, case (ii), performs especially poorly because twice as many qubits are exposed to noise as in the unencoded case and no measures are taken to suppress it.}
	\label{fig:comparison}
\end{figure}

\prsubsec{Energy gap protection}
\label{sub:energy_gap_protection}
\noindent
The EGP approach, introduced in Ref.~\cite{Jordan:2006jb}, appears quite different.  It uses a constant-in-time control Hamiltonian, given by a sum of the stabilizer generators, $\hc^{\textrm{EGP}}(t) = -\alpha \sum_{m=1}^{N_g} S_m$, with $\alpha >0$. States in the codespace are then eigenstates of $\hc$ with eigenvalue $-\alpha N_{g}$, but any state outside the codespace is subjected to an energy penalty.  Since $\hc$ is a function only of the stabilizer generators, $\uc^{\rm EGP}(t)$ again commutes with the code's logical operators which comprise $\haqcb$, so we can write the Hamiltonian in the toggling frame as:
\begin{align}
	\notag
	\tilde H_{\rm EGP}(t) 
		&=  \haqcb + \sum_{j=1}^{N_e}  \tilde E_j^{\rm EGP}(t)\otimes \tilde B_j(t)
\end{align}
Error operators in the EGP toggling frame can be shown to take the form:
\begin{align}
	\tilde E_j^{\rm{EGP}}(t)
	\notag
	&=  E_j e^{\left(2 i \alpha t \sum_{\{S_m,E_j\}=0} S_m \right)}\\
	&=  e^{\left(-2 i \alpha t \sum_{\{S_m,E_j\}=0} S_m \right)} E_j,
	\label{eq:toggling_EGP}
\end{align}
where the sums are taken over all stabilizer generators $S_m$ that anti-commute with the error operator $E_j$.  To obtain this expression we have exploited the following: (i) the stabilizer generators commute with each other (allowing easy manipulation of the exponential operators), and (ii) each generator either commutes or anti-commutes with the noise operators: $S_m E_j = \pm E_j S_m$.  Let $w_j$ be  the number of generators that anticommute with $E_j$.  Then the action of this toggling frame Hamiltonian on any state, $\tilde{\ket{\Psi}}= \tilde{\ket{\psi_c}}\otimes \tilde{\ket{\phi}} \in \mathcal{H}_\textrm{sys} \otimes \mathcal{H}_\textrm{env}$ with $\tilde{\ket{\psi_c}}$ in the codespace is, 
\begin{equation}
\label{eq:heffEGP}
	\tilde H_{\rm {EGP}}(t)\tilde{\ket{ \Psi}} = \left(\haqcb + \sum_{j=1}^{N_e} E_je^{2 i w_j \alpha t}\otimes \tilde B_j(t) \right)\tilde{\ket{ \Psi}}
\end{equation}
Thus the coupling term $E_j\otimes \tilde B_j(t)$ is modulated by a factor of $e^{2 i w_j \alpha t}$. 
Just as in the case of DD (above), the error terms are modulated by an oscillating function in the interaction picture -- which ensure that they average to zero on sufficiently long timescales as long as the frequency of oscillation is larger than the typical frequencies in $\tilde B_j(t)$. In the case of EGP the oscillations are smooth and sinusoidal, whereas for (impulsive) DD the oscillations are square waves in time.  EGP can be made to mimic a decoupling sequence by choosing $\alpha$ so that the EGP oscillations match the frequency of a DD sequence.  Numerical studies provide evidence that in such cases EGP and DD suppress errors equally well, as shown in Fig.~\ref{fig:comparison}. 

In fact, there no requirement that the weights $\alpha$ be constant in time or equal across the stabilizer generators.  Many dynamical decoupling schemes vary the time interval between the pulses; For example, Uhrig's dynamical decoupling scheme (UDD) \cite{Uhrig07} chooses the pulse arrival times as $t_n = T \cos(n \pi/2(N+1))$, where $N$ is the total number of pulses in time interval $\tau$.  To mimic this UDD sequence, where the modulation frequency is not constant in time, we choose a time dependent weight term, $\alpha(t) = NT/\sqrt{t(T-t)}$.  This approach was used, suitably regularized and in the context of a single qubit, by the authors of Ref.~\cite{Jones:2012tr} to produce an effective UDD sequence using continuous controls.  More generally, allowing $\alpha$ to vary in time allows the strongest identification between the DD and EGP approaches, and a unified treatment of both as quantum control protocols. For instance, choosing $\alpha_j(t) = \sum_i \pi \delta(t-t_i^j)/2$ applies $S_j$ at time $t_i^j$ as a unitary operator (impulsive DD), but in the EGP formalism, and points to how one might smoothly interpolate between the two methods.  In this approach, optimal control techniques can be used to choose the $\alpha_j(t)$ to optimally mitigate the system-bath interaction.

\prsubsec{A few important differences}
\label{sub:a_key_difference}
\noindent
The discussion above indicates that DD and EGP suppression methods have very similar behavior with respect to noise and are capable of providing approximately equivalent error suppression.  This relationship is further examined in the context of filter functions in Appendix B and in a dynamical framework in Ref. \cite{Sarovar:2013}. But while these error suppression techniques are closely related, it is important to be aware of some key differences, including the relative ease of their physical implementation and their effect on thermalization.

Many codes possess high-weight stabilizer generators.  These codes cannot reasonably be implemented by EGP, since high-weight Hamiltonians are experimentally infeasible.  They \emph{can} be implemented by DD, however, by exploiting the fact that unitary operators may be generated by many different Hamiltonians.  A high-weight DD pulse can be generated by single-body Hamiltonians, e.g.
\begin{align}
	\notag
	XXX 	&= e^{-i\frac{\pi}{2} XXX}\\
			&= e^{-i\frac{\pi}{2}  \left(XII + IXI + IIX\right)}
\end{align}
However, the single-body Hamiltonian implementing the DD does not commute with the encoded AQC Hamiltonian.  If its implementation is not impulsive (i.e., the DD Hamiltonian is not significantly stronger than the AQC Hamiltonian), then it will be imperfect.  The error can be computed fairly easily with the Baker-Campbell-Hausdorf formula and vanishes in the impulsive limit.  Despite this complication, DD appears to be the only option for implementing high-weight stabilizer generators.  In fact, there exist codes \cite{Ganti2013} that possess a large number of 2-body stabilizer generators and only a few high-weight generators.  These codes may benefit from a hybrid approach where low-weight generators are added as energy penalties, while those of high-weight are included as DD pulses.

Another important difference between DD and EGP is their behavior in thermal environments.  Whereas EGP establishes a real energy difference between the codespace and the various syndrome spaces, DD does not.  If a system with an EGP Hamiltonian is coupled to a cold thermal reservoir, thermalization will lead to a Gibbs distribution ($\rho\propto e^{-\beta H}$), in which the syndrome spaces are thermally populated according to their energy, and the codespace is preferentially populated (since it has the lowest energy).  In contrast, DD creates no real energy difference between syndrome spaces, and the system's steady-state population will be uniformly distributed across all the syndrome spaces.  This distinction between EGP and DD will be important in Sec.~\ref{sec:local_cooling}.



\section{Error correction} 
\label{sec:error_correction}
The error suppression mechanisms described above reduce the rate at which errors appear in the system, by effectively renormalizing the system-bath coupling (see Ref. \cite{Sarovar:2013} for details).  But the coupling cannot be eliminated completely without using an infinite amount of energy (e.g., DD pulses applied at infinitely high frequency, or an infinitely strong EGP Hamiltonian).  Since our resources are finite, physical errors will still accumulate over long timescales, and eventually cause logical errors.  Typical AQC problems (e.g. combinatoric optimization) may have some intrinsic robustness to logical errors (e.g., a few bits flipped in the solution to a Boolean satisfiability problem can be fixed efficiently with classical post-processing), but as they accumulate, they \emph{will} cause the computation to fail.  Using large systems for long computations thus still requires some form of error \emph{correction}.

Implementing error correction requires, to start, utilizing a code that is error-correcting (rather than just error-detecting).  Not only must each error $E_{k}$ anticommute with at least one stabilizer generator, but the set of generators with which it anticommutes ($E_k$'s \emph{syndrome}), must uniquely identify $E_k$'s effect on the encoded computation, and therefore how to correct it.  For simplicity, we will consider \emph{nondegenerate} codes \cite{Smith2007}, in which each correctable error's syndrome is unique
	\footnote{Generalization to degenerate codes simply requires us 
	to interpret the operators $\enu$ as the correction operators 
	associated with syndrome $\bnu$ rather than with errors.}.
Such codes can thus determine \emph{what} error has occurred (instead of simply detecting that \emph{some} error occurred).  Each correctable error can be labeled by its syndrome, a binary vector $\boldsymbol{\nu}$ indicating which stabilizer generators anticommute with the error:
\begin{equation}
\label{eq:nudef}
	S_j \enu S_j \enu = (-1)^{{\boldsymbol{\nu}}_j}.
\end{equation}
For example, in a code with four stabilizer generators, an error labeled $E_{0101}$ commutes with $S_1$ and $S_3$, but anticommutes with $S_2$ and $S_4$.  A code for which the number of correctable errors is exactly equal to the total number of nontrivial syndromes (i.e., $2^{N_g}-1$) is called \emph{perfect} \cite{Got-1997,Gaitan2008}.

\prsubsec{A challenge for error correction}
\noindent
The simplest way to incorporate error correction into an AQC is to just do the encoding, and nothing else.  Perform the encoded adiabatic interpolation, then measure both the stabilizers and the logical $Z$ operators on the final state.  In the absence of errors, the stabilizer measurements would all yield $+1$, and the logical $Z$ measurements would yield the answer to the computation.  If one or more physical errors \emph{do} occur, and accumulate to produce a \emph{correctable} error, then the correct answer can be decoded (classically) from the final measurement outcomes.  So this na\"ive implementation of error correction offers at least one advantage over pure error suppression.

However, it is unlikely to be sufficient.  In the absence of some correcting mechanism \emph{during} the evolution, errors will accumulate.  With a probability that approaches 1 as the computation grows (in size and time), a logical error will occur, and the (decoded) final answer will be wrong.  Thus, we anticipate an additional need for some ongoing entropy-extracting process.

First, however, we must address a different (yet arguably more pernicious) problem, which arises even for very small systems where bath-induced uncorrectable errors are rare: {the encoded problem Hamiltonian transforms correctable errors to logical errors}.

To illustrate this issue, we consider a simple case.  Suppose the system is encoded and initialized in the ground state $\ket{0}$ of an initial Hamiltonian, and evolves unperturbed under the adiabatically changing Hamiltonian until, at time $\tau$, a correctable Pauli error $\enu$ occurs.  Then the system evolves unperturbed through the end of the AQC, at which point we measure the code stabilizers.  Because the Hamiltonian always commutes with the stabilizers and only a single correctable error has occurred, one might think that the error can be detected and identified by its syndrome, $\boldsymbol{\nu}$, 
enabling a restoration of the system to the ground state (in the codespace) by an application of $\enu$.  Unfortunately, in the timespan between the error and its subsequent correction, things go horribly awry.

The overall evolution according to our simplified error model is:
\begin{equation}
\label{eq:evsimp}
	\ket{\psi}_T = \enu \mathcal{U}_\textrm{AQC}(\tau, T) \enu \mathcal{U}_\textrm{AQC}(0,\tau) \ket{0; \bz}_0,
\end{equation}
where the unitary evolution generated by the adiabatic Hamiltonian is given by a time-ordered exponential,
\begin{equation}
\label{eq:uaqc}
\mathcal{U}_\textrm{AQC}(\tau, T) = \exp_+\left(-i\int_{\tau}^T \bar H_{\rm AQC}(s)ds \right).
\end{equation}
(We neglect the as-yet-unspecified error suppressing control Hamiltonian, as its presence does not change the result.)
We assume that the only error that occurs is $\enu$, and that there are not other deviations (such as Landau-Zener transitions) from ideal adiabatic evolution, so $\mathcal{U}_\textrm{AQC}(0,\tau) \ket{0; \bz}_0 = \ket{0; \bz}_{\tau}$.  Now, the encoded AQC Hamiltonian, $\bar H_{\rm AQC}(s)$, is a weighted sum of the code's logical $X, Y$ and $Z$ operators.  Each is a Pauli operator, so it either commutes or anticommutes with the error operator $\enu$.  The encoded AQC Hamiltonian (at any normalized time $s$) splits into a commuting and an anticommuting term, 
\begin{equation}
\label{eq:decomp}
	\bar H_{\rm AQC}(s) = \bar H_{\boldsymbol{\nu}}^+(s)+\bar H_{\boldsymbol{\nu}}^-(s),
\end{equation}
where $[\bar H_{\boldsymbol{\nu}}^+(s), \enu] = \{\bar H_{\boldsymbol{\nu}}^-(s), \enu\} = 0$.  After some algebra, Eq.~\eqref{eq:evsimp} becomes
\begin{equation}
\label{eq:evsimp2}
		\ket{\psi}_T = \exp_+\left(-i\int_{\tau}^T \left( \bar H_{\boldsymbol{\nu}}^+(s)-\bar H_{\boldsymbol{\nu}}^-(s)\right)ds \right) \ket{0; \bz}_\tau
\end{equation}
Between the time when the error happens ($\tau$) and when it is corrected ($T$), the encoded system experiences a new, \emph{effective} Hamiltonian, $\bar H^\prime_\bnu = \bar H_{\boldsymbol{\nu}}^+(s)-\bar H_{\boldsymbol{\nu}}^-(s)$.  Since the state $\ket{0; \bz}_\tau$ is not generally an eigenstate of $\bar H^\prime_\bnu$, the system will undergo unintended evolution within the codespace, moving it out of the ground state (a logical error).  

The root problem here is that the encoded adiabatic Hamiltonian acts differently on different syndrome spaces (eigenspaces of the stabilizers).  Correctable errors flip various stabilizers' eigenvalues, moving the system from one syndrome space to another.  Only on the codespace itself is $\bar{H}_{\rm AQC}$ guaranteed to act like the original problem Hamiltonian $H_{\rm AQC}$.  In other words, the AQC Hamiltonian itself rapidly turns correctable errors into logical errors.  Like non-transversal implementations of logic gates in the circuit model, na\"ively encoded AQC Hamiltonians cause errors to propagate, and decoding at the end of the computation is unlikely to be effective.

The problem persists even if errors are corrected during the computation.  Even the best error correction is not instantaneous, so errors will survive for some time before being corrected.  In large computations, the equilibrium between noise and correction occurs at a finite density of errors, meaning that at every instant the system will be out of the codespace.  Thus, it is absolutely necessary to modify the encoded Hamiltonian so that $\bar{H}_{\rm AQC}$ is sufficiently similar to $H_{\rm AQC}$ not just on the codespace, but on every likely syndrome space.
 
\subsection{Protected Hamiltonians}
\label{sub:protected_hamiltonians}
The promotion of physical errors to logical errors by the adiabatic Hamiltonian poses a serious threat to error correction in AQC using stabilizer codes.  In this section, however, we demonstrate that this threat may be avoided by carefully choosing the logical operators -- but doing so comes at a steep cost: these \emph{protected} logical operators are complicated sums of high-weight Pauli operators.

Physical errors become logical errors because $\bar{H}_{\rm AQC}$ does not act identically on all of the syndrome spaces.  In principle, it would be enough if any error mapped the \emph{ground} state[s] of $\bar{H}_{\rm AQC}$ to another eigenstate of $\bar{H}_{\rm AQC}$,
$$
\bar H_{\rm AQC}(t) \enu \ket{0,k;\bz}_t = \epsilon_{0,k;\bnu}(t) \enu \ket{n,k;\bz}_t.
$$
But our error correcting code is not allowed to ``know'' what the ground state is.  This condition has to hold for any encoded logical Hamiltonian built as a sum of logical operators.  
So, whenever an AQC is encoded in a stabilizer code, some logical Hamiltonians will rapidly transform some physical errors into logical errors unless the following stronger condition is satisfied: {for any error $\enu$ and any eigenstate $\ket{n,k}$ of the logical Hamiltonian, $\enu \ket{n,k;\bz}$ must be an eigenstate of the encoded logical Hamiltonian.}  In other words,
\begin{equation}
\label{eq:eigentoeigen}
	\bar H_{\rm AQC}(t) \enu \ket{0,k;\bz}_t = \epsilon_{n,k;\bnu}(t) \enu \ket{0,k;\bz}_t
\end{equation}
If this condition holds, then at least adiabatic evolution is still possible in the presence of errors (although errors might reduce or eliminate the gap, or even cause an excited state to become the ground state).  

If we want Eq.~\eqref{eq:eigentoeigen} to hold, then the encoded Hamiltonian (and therefore the encoded logical operators) is constrained.  The eigenstates specified in Eq.~\eqref{eq:eigentoeigen} form a complete orthonormal basis, so the most general encoded Hamiltonian that satisfies Eq.~\eqref{eq:eigentoeigen} is a sum of their projectors,
\begin{align}
\label{eq:mostmostgeneral}
	\bar H_{\rm AQC}(t) 
		\notag
		&= \sum_{n,k,\bnu} \epsilon_{n,k;\bnu}(t)\ket{n,k;\bnu}_t\!\bra{n,k;\bnu}\\
		&= \sum_{n,k,\bnu} \epsilon_{n,k;\bnu}(t)\enu\ket{n,k;\bz}_t\!\bra{n,k;\bz}\enu^\dagger.
\end{align}
The projector $\ket{n,k;\bz}_t\!\bra{n,k;\bz}$ may be written as a product of the projector onto the codespace (expressible entirely in terms of code's stabilizers) and the logical projector onto the encoded state $\ket{n,k}_t$ (expressible entirely in terms of the code's logical operators).   

To construct an encoded Hamiltonian with these properties, we can exploit the freedom in defining logical operators of a code. Multiplying a logical operator by a stabilizer operator yields an equally valid representation of the same logical operator.  We can represent elements of the stabilizer group by binary vectors $\mathbf{j}$, of length $N_g$, in which a $1$ indicates the presence of the corresponding generator, i.e. $S_{\mathbf{j}} = S_1^{j_1}S_2^{j_2}\cdots S_{N_g}^{j_{N_g}}$.  For example, if there are four stabilizer generators, then $S_{0110} \equiv S_2 S_3$.  In this notation, the fact that (Pauli) errors either commute or anticommute with each of the stabilizers can be written as:
\begin{equation}
\label{eq:escom}
	S_{\mathbf{j}} E_\bnu =  (-1)^{\mathbf{j}\cdot\bnu} E_\bnu S_{\mathbf{j}}
\end{equation}
Then if $\bar L_i$ is a particular encoded logical operator, each stabilizer operator $\Sj$ defines an \emph{equivalent logical operator} $\bar L_i \Sj$ that acts identically on the codespace.  Linear combinations, e.g. $\sum_\mathbf{j}{\beta_{i\mathbf{j}}\bar L_i \Sj}$, are also valid logical operators.  

Now, any problem Hamiltonian can be written as a sum of logical operators,
$$H_{\rm AQC}(t) = \sum_i \alpha_i(t) L_i.$$
When we encode it, we can choose any encoding $L_i \to \sum_\mathbf{j}{\beta_{i\mathbf{j}}(t)\bar L_i \Sj}$ (where the $\bj$ sum is taken over all binary vectors of length $N_g$, and $\bar L_i$ is an arbitrary encoded representation of $L_i$) as long as $\sum_\bj \beta_{i\bj}(t) = 1$.  So, an equivalent encoded problem Hamiltonian is
\begin{equation}
\label{eq:hamform}
	\bar H_{\rm AQC}(t) = \sum_i \alpha_i(t) \bar L_i \sum_\bj \beta_{i\bj}(t) S_\bj,
\end{equation}
Now, we impose the constraint that it satisfy Eq.~\eqref{eq:eigentoeigen}.  Inserting Eq.~\eqref{eq:hamform} into Eq.~\eqref{eq:eigentoeigen} and multiplying by $\enu$ on the left, we obtain
\begin{equation}
\sum_{i} \alpha_i(t) \bar L_i \left( \sum_\bj \beta_{i\bj}(t) \Gamma_{i\bj}^{\bnu}\right) \ket{n,k;\bz}_t = \epsilon_{n,k;\bnu}(t) \ket{n,k;\bz}_t, \nonumber
\end{equation}
where $\enu \bar L_i = \gamma_{i\bnu}\bar L_i \enu$, and $\Gamma_{i\bj}^\bnu = \gamma_{i\bnu}(-1)^{\bj\cdot\bnu}$ is 1 if $\enu$ commutes with $\bar L_i S_\bj$ and $-1$ otherwise.  Direct computation then shows that $\Gamma_{i\bj}^\bnu$ obeys the orthogonality condition
\begin{equation}
\label{eq:gamma}
	\sum_\bnu \Gamma_{i\bj}^\bnu \Gamma_{i\bk}^{\bnu} = 2^{N_g} \delta_{\bj\bk}, ~~~~ \forall i.
\end{equation}

Recall that $\ket{n,k;\bz}_t$ is defined to be an eigenstate of $\sum_i \alpha_i(t) \bar L_i$ for any choice of $\alpha_i(t)$.  Therefore, the sum $\sum_\bj \beta_{i\bj}(t)\Gamma_{i\bj}^\bnu$ cannot depend on $i$ and we may substitute
\begin{equation}
\label{eq:deflambda}
	\lambda_\bnu(t) = \sum_\bj \beta_{i\bj}(t)\Gamma_{i\bj}^\bnu
\end{equation}
With this definition, we see that $\epsilon_{n,k;\bnu}(t)=\epsilon_{n,k;\bz}(t)\lambda_\bnu(t)$.  This means that an encoded operator satisfying Eq.~\eqref{eq:eigentoeigen} acts identically (up to a scale factor $\lambda_\nu$) on each syndrome subspace, including the codespace.  Correctable errors map eigenstates of $\bar{H}_{\rm AQC}$ in the codespace to eigenstates in the appropriate syndrome space.  In particular, the ground state maps either to the lowest-energy eigenstate in the syndrome space (if $\lambda_\bnu$ is positive) or the highest-energy eigenstate (if $\lambda_\bnu$ is negative). We refer to $\lambda_\bnu$ as a \emph{deformation} factor in Ref. \cite{Sarovar:2013} since it represents the deformation of the codespace energy structure when reproduced within syndrome spaces.

This means that the most general Hamiltonian satisfying condition Eq.~\eqref{eq:eigentoeigen} can be constructed by symmetrizing over errors $\enu$:
\begin{equation}
\label{eq:mostgen}
	\bar H^{\lambda}_{\rm AQC}(t) = \sum_{\bnu} \lambda_\bnu(t) \enu \bar H(t) \projzero \enu
\end{equation}
where $\bar H(t)$ is a Hamiltonian encoded using any logical operators.  We refer to such Hamiltonians as \emph{protected Hamiltonians} since adiabatic evolution under them does not lead to propagation of errors in the correctable syndrome subspaces. Furthermore, we define the \emph{canonical} protected Hamiltonian by setting all $\lambda_\bnu$ in Eq.~\eqref{eq:mostgen} to 1,
\begin{equation}
\label{eq:proham}
	\bar H^{\rm p}_{\rm AQC}(t) = \sum_{\bnu} \enu \bar H(t) \projzero \enu.
\end{equation}
The canonical protected Hamiltonian acts truly identically on every syndrome space, with no scale factor at all.  Similar Hamiltonians were introduced in a slightly different context in Ref.~\cite{Jacobs2013}.

Unfortunately, implementing the canonical protected logical operators seems infeasible.  They generally contain (many) terms of very high weight, up to the total number of qubits in the system.  However, if we allow a scale factor (i.e., choose non-canonical but still protected logical operators), then it is possible to reduce the maximum weight by choosing the coefficients $\lambda_\bnu$ to eliminate the highest weight logical operators from Eq.~\eqref{eq:mostgen}.  Inverting Eq.~\eqref{eq:deflambda}, we have,
\begin{equation}
\label{eq:defbij}
	\beta_{i{\mathbf{j}}}(t) = \sum_\bnu \frac{1}{2^{N_g}}\lambda_\bnu(t) \Gamma_{i\bj}^\bnu,
\end{equation}
which can be used to choose the $\lambda_\bnu(t)$'s such that the $\beta_{i\bj}(t)$'s corresponding to high weight operators vanish.  Note that we have explicitly included the time dependence in $\lambda_\bnu(t)$ and $\beta_{i\bj}(t)$ to maintain generality, but it may be convenient to force them to be time \emph{independent}, as in the canonical protected Hamiltonian (where $\lambda_\bnu(t)=1$).  The time dependence may be useful for constructing more sophisticated error suppression schemes, such as increasing the penalty if a particular error is otherwise more likely at a certain time, but we do not consider such schemes here.

In the appendix, we examine the $[[5,1,3]]$ code \cite{Got-1997} and show that while the canonical protected logical operators are weight-5, we can construct protected logical operators with weight 3.  \emph{Any} logical operator for a distance-$d$ code must have weight at least $d$, so it is encouraging that protected operators of weight $d$ exist in this case.  If such a construction is possible for any code (an open question), it would remove one obstacle to implementation.

However, a further challenge is the sheer number of Pauli operators terms required to implement these protected logical operators; for example, the canonical protected logical operators for $n$-qubit codes are sums of $O(2^n)$ Pauli operators.  The minimum-weight protected logical operators that we construct in the appendix are simpler, but still contain $O(2^d)$ distinct Paulis.  Applying a single weight-$d$ operator (for $d\gg1$) as a Hamiltonian is already challenging (see concluding discussion).  Balancing many such terms seems fantastically difficult.  We suspect it will be feasible \emph{only} if there exist protected logical operators in which the required sum of Pauli operators can be factored or otherwise decomposed into a sum of a few products.  For example, $(X+Z)^{\otimes n}$, expanded as a sum of Paulis, contains $2^n$ terms -- but because it factors, it is no harder to implement than $X^{\otimes n}$.  We are not aware of any such structure in protected logical operators, but further research might reveal one.


\section{Error correction by local cooling} 
\label{sec:local_cooling}

Suppose that a way is found to prevent the adiabatic Hamiltonian from converting physical errors to logical errors (e.g., by implementing protected logical Hamiltonians).  AQC would then face ``only'' the same problem that confronts circuit-model computation; errors accumulate over time.  These errors are the manifestation of entropy injected into the system by coupling to the bath, and unless we actively pump that entropy out, the computation is likely to fail within a relatively short time.  Active error correction, however, requires fast gates and high-weight stabilizer measurements during the computation -- both of which are outside the standard AQC toolset (and infeasible in many of the physical systems on which AQC might be implemented).  \emph{Local cooling} offers a potential route around this difficulty.

In the local cooling model, each physical qubit is coupled to a very low temperature bath that serves as the entropy sink for the system.  The coupling is designed such that if the stabilizer generators are added to the Hamiltonian, the bath is able to detect the increase in energy associated with an error and then to absorb that energy and reverse the error.  The dynamics of error correction by cooling is worked out in detail in Ref. \cite{Sarovar:2013} and we sketch the scheme here. 

We will consider the Hamiltonian of Eq.~\eqref{eq:aqchambar} with an additional local coupling of each error operator to a cold, damped reservoir given by 
\[\hr=\sum_j E_j \otimes \sum_k g^{(j)}_k (b_k^{(j)} + b^{\dagger (j)}_k) ).\]  
In addition to this interaction Hamiltonian, the reservoirs have free Hamiltonians of independent oscillators: $H_{\rm R} = \sum_j \sum_k \omega^{(j)}_k b^{\dagger (j)}_k b^{(j)}_k$. We assume that each reservoir has a broad frequency distribution and is well damped such that it is in its ground state at all times with high probability.  It is shown in Ref. \cite{Sarovar:2013} that such a system-reservoir coupling, a control Hamiltonian of the EGP type and a protected Hamiltonian implementation of the logical Hamiltonian are sufficient for an automated implementation of error correction where the excitations induced by local Pauli perturbations from the bath are quenched by the cold reservoir.

However, because the reservoir is coupled to the system through low weight Pauli operators ($E_j$), it can only cool local excitations.  Cooling away a high-weight error would have to be accomplished through a sequence of single-Pauli operations that reduce the error weight until the system is returned to the codespace. However, as shown in Ref. \cite{Sarovar:2013} the cooling dynamics is a biased random walk in syndrome space and therefore in order for such a sequence of single Pauli operations to successively cool away several errors the energy landscape of the system must be such that the energy penalty associated with an error increases with its weight. That is, the energy of states in a correctable syndrome subspace must increase monotonically with the weight of the error that takes the codespace to that syndrome space. 
The EGP control Hamiltonians corresponding to most stabilizer codes do not have this property (we shall discuss exceptions to this in a moment). For example, for the abelian toric code \cite{Kit-2003}, a state with two errors can have the same energy penalty as a state with one error if the two errors are neighboring, \textit{i.e.} a string excitation does not have an excited state energy proportional to the string length and a local measurement of energy cannot distinguish between such degenerate errors. Such models have no ``string tension'' \cite{Kay2008}. 

EGP control Hamiltonians that provide a favorable landscape, where syndrome subspace energies scale with error weight, may be constructed explicitly as
\begin{equation}
\label{eq:nicelandscape}
	H_{\rm C}^{\rm EGP}  = \sum_{k} \sum_{\substack{\enu\;\rm{s.t.} \\ \mathbf{w}(\enu)=k}} \delta(k) \enu \projzero \enu
\end{equation}
where $\delta(k)>0$ is the energy penalty associated with weight $k$ errors and $\delta(k)<\delta(k+1)$. However, as in the case of the protected Hamiltonian for logical evolution, such constructions result in Hamiltonians that are very high weight. In fact, recent work has shown that there are significant obstacles to constructing systems with the kind of energy landscape discussed above, known as a \emph{self-correcting quantum memory}, using \emph{local} stabilizers in two dimensions \cite{Bravyi:2009kp,LandonPoulin2013}.  These results do not, however, rule out construction of an energy landscape for which low weight operators are penalized according to their weight, but higher weight operators are not.  Such a code would exhibit string tension that ``snaps'' after the string length grows too long, and would likely provide enhanced protection over standard codes.  It might be constructed as in Eq.~\eqref{eq:nicelandscape} by choosing the energy penalties so that high-weight stabilizers cancel, as we did in Section \ref{sub:protected_hamiltonians} to construct the protected logical operators.  
Furthermore, although the restriction to two-dimensional (planar) codes may be appealing from an engineering perspective, self-correcting lattice codes in four (and perhaps three) spatial dimensions do exist\cite{Haah2011,Bravyi2011,Bravyi2011b,Michnicki2012,Dennis2002}.  Embedding such higher-dimensional codes into two dimensions requires non-planar connectivity, but this is not necessarily unrealistic; superconducting qubit systems routinely construct non-planar interaction graphs by coupling distant qubits with wires \cite{dwave}. 

\prsec{Discussion}
\noindent
This work arose from our attempts to answer the question, ``Can AQC be made fault tolerant?''  We began without a clear picture of what `fault tolerance' should mean in the adiabatic context.  Taking cues from the theory of \emph{circuit model} fault tolerance, we believed that developing a clear understanding of error suppression and error correction in AQC would be a necessary first step, whatever the ultimate definition turned out to be.  In this paper, we have presented such a framework, investigating relationships between error suppression methods and describing a serious challenge to adiabatic error correction.  Through the methods of protected Hamiltonians and local cooling, we have even suggested techniques for avoiding these obstacles and correcting errors without resort to circuit model gates or syndrome measurements.

Our analysis, however, falls short of establishing a threshold theorem for adiabatic fault tolerance. We cannot prove fault tolerance because controlling the encoded AQC (using slowly varying Hamiltonians rather than fast gates) seems to require high-weight Hamiltonians. These Hamiltonians are not available in any feasible technology to our knowledge. We are therefore unable to propose a feasible control protocol, and so have no credible model to describe control errors. Without a plausible error model, we cannot attempt to prove fault tolerance or calculate a threshold.  \emph{If} in the future the implementation of protected logical Hamiltonians is shown to be practical, then it may become possible to construct fault tolerant logical-operation protocols, and to prove the existence of a threshold.  But, absent such a breakthrough, logical Hamiltonians on encoded qubits appear to require unphysical resources, leaving us pessimistic that any form of fault tolerance will ever be achieved in a purely adiabatic model of quantum computation.

This conclusion rests on several assumptions.  We now list these assumptions, and the limitations in our analysis, in the hope that future work may find ways to circumvent the obstacles identified here.
 
First, we have assumed that protecting AQC from noise will require the use of high-distance stabilizer codes. This assumption is motivated by fault tolerance in the circuit model, which depends critically on the use of high-distance quantum codes, because the computation must be robust to a constant (albeit low) density of errors.  In the adiabatic setting, high-distance codes lead directly to two specific problems: (i) the encoded logical Hamiltonian quickly transforms correctable errors into uncorrectable logical errors; and (ii) the logical operators that comprise the encoded Hamiltonian necessarily have high weight.  Our work suggests a solution to problem (i), but problem (ii) poses a greater challenge.  In the circuit model, high-weight logical operators are implemented by performing many one- or two-qubit unitaries in parallel.  Though such sequences are more complicated than unencoded gate operations, the (linear in $N$) increase in gate complexity is far outweighed by enhanced (exponential in $N$) resilience against noise.  In the adiabatic model, however, logical operations are implemented as Hamiltonians -- not unitaries.  And whereas high-weight unitaries can be implemented in $O(1)$ time using parallel gates, there is no comparable way to apply high-weight Hamiltonians.  Solving this issue will likely require significant advances.  Perturbative gadgets can approximate high-weight Hamiltonians with only weight-two interactions, but introduce unprotected gauge qubits to the system.  Furthermore, the published analyses \cite{Jordan2008} of gadget perturbation theory require couplings that scale exponentially with the operator weight, though perhaps more sophisticated gadget perturbation theories can be developed which reduce this penalty.  Even more desirable (though correspondingly less likely) would be the development of qubits whose dominant interactions are naturally high-weight \footnote{For example, in gases the dominant thermalization mechanism is three-body collisions.}. Should this obstacle be overcome, our construction of protected logical operators (Sec.~\ref{sub:protected_hamiltonians}) will be highly relevant.

Second, we have assumed that the only available error correction mechanism is \emph{local} cooling. The local nature of the cooling is clearly physically motivated, but local cooling can only drive single qubits and act upon local information \cite{Sarovar:2013}. It is possible that a more sophisticated cooling mechanism that acts on multiple qubits in a neighborhood (similar to a continuous-time   error decoding algorithm) could be constructed.  Such a mechanism might obviate the need for a monotonic energy landscape as stated in Sec. \ref{sec:local_cooling}.  However, we have no constructive ideas for implementing such a cooling mechanism at this time.
 
Finally, an overriding assumption in this whole work is that elements of the circuit model, like fast gates and measurements, are not available. Of course, one could begin incorporating more elements of the circuit model in order to implement error correction. However, in that case, the model of computation begins to look more and more like the circuit model itself.  If fault-tolerant AQC demands development of all the resources required for computation in the circuit model, why bother with AQC at all?

\prsec{Acknowledgements} 
\label{sec:acknowledgements}
\noindent
We acknowledge important discussions with Sandia's AQUARIUS Architecture team, especially with Andrew Landahl and Anand Ganti. This work was supported by the Laboratory Directed Research and Development program at Sandia National Laboratories. Sandia is a multi-program laboratory managed and operated  by Sandia Corporation, a wholly owned subsidiary of Lockheed Martin Corporation, for the United States Department of Energy's National Nuclear Security Administration under contract DE-AC04-94AL85000.
\bibliography{aqc_error_suppression_long}

\begin{appendix}

\section{Subsystem structure of protected Hamiltonian} 
\label{sec:subsystem_structure_of_protected_hamiltonian}

The protected Hamiltonians defined in Eq.~\eqref{eq:mostgen} naturally induce tensor product decomposition on the system Hilbert space into a logical subsystem, $\hilbert_{\rm log}$, and a syndrome subsystem, $\hilbert_{\rm synd}$.  
\begin{equation}
\label{eq:subsystem}
	\bar \hilbert_{\rm sys} = \left(\hilbert_{\rm log} \otimes \hilbert_{\rm synd}\right)\oplus \hilbert_{\rm uncorr}.
\end{equation} 
Correctable eigenstates of the system, $\ket{n,k;\bnu}_t$ may be written in this context as $\ket{n,k}\otimes\ket{\bnu}_t$.  If the Hamiltonian is encoded as in Eq.~\eqref{eq:proham}, then its action on any correctable state is
\begin{equation}
\label{eq:hamaction}
	\bar H_{\rm AQC}^{\rm p}(t) \ket{n,k}\otimes\ket{\nu}_t = \epsilon_{n,k} \ket{n,k}\otimes\ket{\nu}_t
\end{equation}
Thus the protected logical operators themselves inherit this tensor product structure when acting on correctable states:
\begin{equation}
	\bar L^{\rm p} = \sum_{\bnu} \enu \bar L \projzero \enu \rightarrow L\otimes \mathbf{I}_{n-k},
\end{equation}
where $\bar L$ is any logical operator of the code, and $\mathbf{I}_{n-k}$ is the identity operator on the $2^{n-k}$ dimensional syndrome space.  Assuming a perfect code, a set of operators may be constructed that act as Pauli operators on the syndrome bits.  The stabilizer generators of the code may be interpreted as the Pauli $Z $operators on the syndrome bits:
\begin{equation}
\label{eq:paulis}
	S_j \rightarrow I\otimes Z_j
\end{equation}
Pauli $X$ operators flip stabilizer bits without introducing a phase, and so may be constructed as
\begin{equation}
\label{eq:stabx}
	\sum_{\boldsymbol{\nu}} E_{\boldsymbol{\nu}\oplus \mathbf{j}} \mathbf{P}_0 \enu \rightarrow I\otimes X_j
\end{equation}
And the Pauli $Y$ operators can be constructed using the Pauli relation, $iXZ = Y$:
\begin{equation}
\label{eq:staby}
	i \sum_{\boldsymbol{\nu}} E_{\boldsymbol{\nu} \oplus \mathbf{j}} \mathbf{P}_0 \enu S_j \rightarrow I\otimes Y_i
\end{equation}

These operators now allow us to represent any physical Pauli operator  in terms of its action on the logical and syndrome degrees of freedom.  Any logical operator of the code, $L$, may be written as $L\rightarrow L\otimes O$ for some operator, $O$ that acts entirely on the syndrome space.  Because the protected operators take the form $L\otimes I$, we may determine $O$ by multiplication:
\begin{equation}
\label{eq:ooperator}
	L^{\rm p} L \rightarrow (L\otimes I) (L\otimes O) = I\otimes O	
\end{equation}
Using this decomposition, one may easily see why sums of logical operators do not necessarily act on syndrome subspaces in the same way as they do on the codespace: their associated syndrome operator is different. 

\prsubsec{Example: $[[5,1,3]]$ code}
\label{sub:example_with_5_1_3_code}
\noindent
Consider the quantum $[[5,1,3]]$ code, defined in terms of the stabilizer generators,
\begin{align}
	S_1 &= IXZZX \\
	S_2 &= XIXZZ \\
	S_3 &= ZXIXZ \\
	S_4 &= ZZXIX
\end{align}
and logical operators
\begin{align}
	\bar X &= XXXXX\\
	\bar Z &= ZZZZZ
\end{align}
In the tensor product basis defined above, these logical operators take the form:
\begin{align}
	\notag
	 \bar X \rightarrow	
		 X\otimes \frac{1}{4}
		&\left(  -  IIII +  IIIZ +  IIZI +  IIZZ \right.\\ \notag
		 &	+  IZII -  IZIZ +  IZZI +  IZZZ \\ \notag
		 &  +  ZIII -  ZIIZ -  ZIZI -  ZIZZ \\ \notag
		& \left.	+  ZZII -  ZZIZ +  ZZZI +  ZZZZ\right) \\
	\notag
 	\bar Z \rightarrow	
		 Z \otimes \frac{1}{4}
		&\left( - IIII + IIIZ + IIZI - IIZZ \right.\\ \notag
		 & + IZII + IZIZ - IZZI - IZZZ \\ \notag
		 & + ZIII + ZIIZ + ZIZI + ZIZZ \\ \notag 
		 &\left. - ZZII + ZZIZ - ZZZI + ZZZZ\right)
\end{align}
The syndrome part of these operators are different, and thus will behave differently on each syndrome subspace.  To correct this we introduce the protected operators, defined by Eq.~\eqref{eq:proham}, which in this case take the form,
\begin{align}
	\notag
	 \bar X^{\rm p} = 
		 &- IIZXZ - IXIYY - IYYIX - IZXZI -\\ \notag
		 &XIYYI - XXXXX - XYZZY - XZIIZ -\\ \notag
		 &YIXIY - YXYZZ - YYIXI - YZZYX -\\ \notag
		 &ZIIZX - ZXZII - ZYXYZ - ZZYXY
 \\
	\notag
	 \bar Z^{\rm p} = 
		&- IIYZY - IXXIZ - IYZYI - IZIXX -\\ \notag
		& XIZIX - XXIZI - XYYXZ - XZXYY - \\ \notag
		& YIIYZ - YXZXY - YYXZX - YZYII - \\ \notag
		& ZIXXI - ZXYYX - ZYIIY - ZZZZZ
\end{align}
These operators in the tensor product basis are simply $X^{\rm p}\rightarrow X\otimes I$ and $Z^{\rm p}\rightarrow Z\otimes I$.  These operators act consistently, but require the use of high-weight operators.  However, if we add up the \emph{minimum weight} versions of each logical operator, we have 
\begin{align}
	\notag
	 \bar X_3 =
		& - IIZXZ - IXIYY - IYYIX - IZXZI - \\ \notag
		& XIYYI -  XZIIZ - YIXIY - YYIXI - \\ \notag
		& ZIIZX - ZXZII \\ \notag
	 \rightarrow &X \otimes \left(2\proj_0 + \mathbf{I}_4/2\right)
 \\
	\notag
	\bar Z_3 =
		& - IIZXZ - IXIYY - IYYIX - IZXZI - \\\notag
			& XIYYI - XZIIZ - YIXIY - YYIXI  \\\notag
			& - ZIIZX - ZXZII \\\notag
	\rightarrow & Z \otimes \left(2\proj_0 + \mathbf{I}_4/2\right)
\end{align}
Because the syndrome part of these operators is the same, these operators act consistently across subspaces, avoiding the problems we described above.  It remains an open question whether such low-weight constructions exist for higher-distance codes.  


\prsec{Generalized filter functions for single qubit case}
\noindent
The relationship between DD and EGP is nicely illustrated by a single qubit example.  We shall consider a single qubit evolution in the presence of pure dephasing noise.  Note that for a single qubit, the error detection code allows us to stabilize a single state ($\ket{+}$ or $\ket{-}$ in the example below), rather than a subspace. The Hamiltonian describing noisy qubit evolution is
	\begin{equation}
	\label{eq:single_ham}
		H(t) = \frac{1}{2} c(t) \sigma_x + \frac{1}{2} \eta(t) \sigma_z.
	\end{equation}
	\begin{figure}[!tr]
		\begin{center}
		\includegraphics[width=\columnwidth]{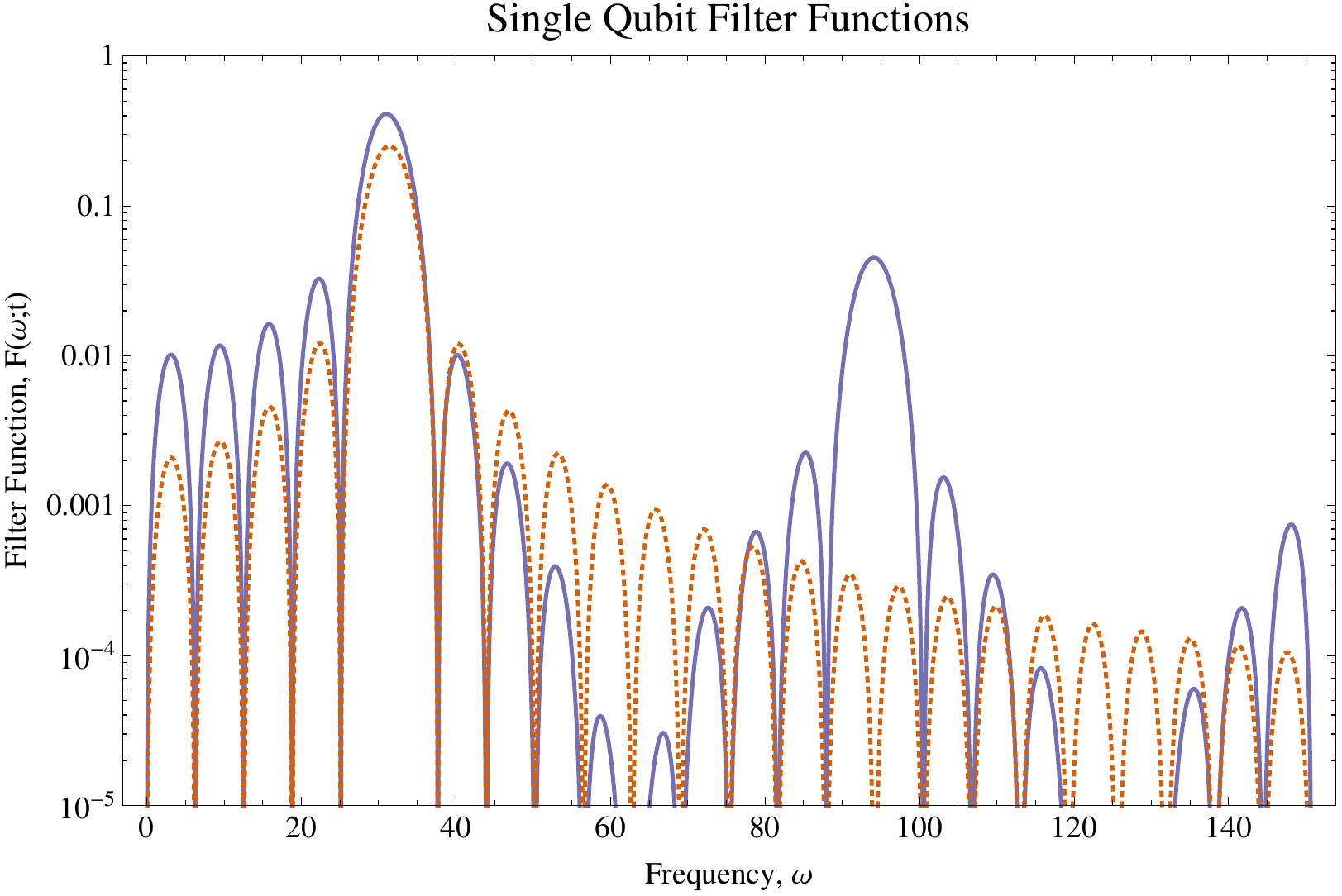}
		\caption{The filter functions for periodic dynamical decoupling (solid) and continuous amplitude driving (dashed).  The filter minima can be made to overlap at first order in the Magnus expansion.}
		\label{fig:filter_fns}
		\end{center}
	\end{figure}
Here $c(t)$ is the control field that could either implement DD or EGP, and $\eta(t)$ is the stochastic noise. Dynamical decoupling proceeds by applying a sequence of X-type pulses to perturbatively decouple the noise, while EGP introduces a time-independent energy penalty for noise-induced error transitions. We consider a classical approximation of the system-bath coupling and represent the bath fluctuations as a classical stochastic process. This is not necessary for the following but we do so for simplicity. 
	In the toggling frame, this Hamiltonian takes the form,
	\begin{equation}
	\label{eq:single_rot}
		\tilde H(t) = \frac{1}{2} \eta(t) \sigma_z \exp(-2i\int_0^t c(s) \,ds\, \sigma_x)
	\end{equation}
	Defining the fidelity as the probability of finding the state in the $\ket{+}$ state after the evolution,
	\[ \mathcal{F}(\tau) = \expect{\abs{\bra{+} U(\tau) \ket{+}}^2}_\eta\]
	where $\expect{\cdot}_\eta$ indicates a classical stochastic average over instances of the noise, $\eta(t)$.  To first order in the Magnus expansion \cite{Oteo00}, the effective unitary operator is
	\begin{equation*}
		U(t) \simeq \exp\left[\frac{-i}{2} \int_0^\tau ds\, \eta(t)\left(\cos(\chi(s))\sigma_z + \sin(\chi(s))\sigma_y \right))\right]
	\end{equation*}
	Where $\chi(t)= \int_0^t c(s) ds$.  The fidelity is then,
	\begin{widetext}
	\begin{align*}
	\label{eq:inner_prod_x}
		\mathcal{F}(\tau) 
			&= \expect{\frac{1}{2}\cos^2\left(\int_0^\tau ds\, \eta(t)\left(\cos(\chi(s))\sigma_z + \sin(\chi(s))\sigma_y \right)\right)}_\eta\\
			&= \frac{1}{2}+\frac{1}{2}\exp\left(-\frac{1}{2}\int_0^t ds_1 \int_0^{s_1} ds_2\expect{\eta(s_1)\eta(s_2)}_\eta\cos(\chi(s_1)-\chi(s_2))\right) \\
			&= \frac{1}{2}+\frac{1}{2}\exp\left(-\int_{-\infty}^\infty d\omega \,S(\omega)F(\omega)\right)
	\end{align*}
	This defines the filter function as 
	\begin{equation*}
	\label{eq:filterfn}
		F(\omega;\tau) = \frac{1}{\omega^2}\int_0^\tau ds_1 \int_0^\tau ds_1 \cos(\chi(s_1)-\chi(s_2))) \cos((s_1-s_2)\omega)
	\end{equation*}
	\end{widetext}
	For EGP, $\chi(t) = \omega t$, while for dynamical decoupling $\chi(t) = n_t\pi$, where $n_t$ is the number of pi-pulses applied up to time $t$.  The resulting filter functions are shown in Fig.~\ref{fig:filter_fns}. Note that for dynamical decoupling with hard pulses, the effective Hamiltonian converges at first order in the Magnus expansion.  Higher order terms will appear for the continuous driving case, but will be negligible for weak noise and short time.

\end{appendix}

\end{document}